\documentstyle[12pt]{article}

\newcommand{\be}{\begin{equation}}
\newcommand{\ee}{\end{equation}}
\newcommand{\bea}{\begin{eqnarray}}
\newcommand{\eea}{\end{eqnarray}}
\newcommand{\nn}{\nonumber \\}
\newcommand{\p}[1]{(\ref{#1})}
\newcommand{\ba}{\begin{array}}
\newcommand{\ea}{\end{array}}
\newcommand{\vs}[1]{\vspace{#1 mm}}

\renewcommand{\a}{\alpha}

\def\bbox{{\,\lower0.9pt\vbox{\hrule \hbox{\vrule height 0.2 cm
\hskip 0.2 cm \vrule height 0.2 cm}\hrule}\,}}
\newcommand{\dsl}{\pa \kern-0.5em /}

\newcommand{\pa}{\partial}

\def\ua{{\underline a}}
\def\ub{{\underline b}}
\def\uc{{\underline c}}
\def\ud{{\underline d}}

\def\a{\alpha}
\def\Th{\Theta}

\def\ua{{\underline a}}
\def\ub{{\underline b}}
\def\uc{{\underline c}}
\def\ud{{\underline d}}
\def\ue{{\underline e}}
\def\uf{{\underline f}}

\begin{document}

\topmargin 0pt
\oddsidemargin 5mm

\renewcommand{\thefootnote}{\fnsymbol{footnote}}
\begin{titlepage}

\setcounter{page}{0}
\begin{flushright}
UG-11/98\\
hep-th/9805065
\end{flushright}

\vs{5}
\begin{center}
{\Large THE M5-BRANE HAMILTONIAN}
\vs{10}

{\large
E. Bergshoeff$^1$, D. Sorokin$^2$\footnote{Alexander von Humboldt fellow. On
leave from Kharkov Institute of Physics and Technology,  Kharkov, 310108,
Ukraine.} and  P.K. Townsend$^3$\footnote{On leave from DAMTP,  University of
Cambridge, U.K.} } \\
\vs{5}
${}^1${\em Institute for Theoretical Physics, Nijenborgh 4,\\
9747 AG Groningen, The Netherlands.}\\
\vs{5}
${}^2${\em Humboldt-Universit{\"a}t zu Berlin,\\
Mathematisch-Naturwissenschaftliche Fakultat,\\
Institut f{\"u}r Physik, Invalidenstrasse 110,\\
D-10115 Berlin, Germany.}\\
\vs{5}
${}^3${\em Institute for Theoretical Physics,\\
University of California at Santa Barbara,\\ 
CA 93106, USA.}
\end{center}
\vs{10}
\centerline{{\bf Abstract}}

We obtain the Hamiltonian form of the worldvolume action for the M5-brane in
a general D=11 supergravity background. We use this result to obtain a new
version of the covariant M5-brane Lagrangian in which the tension appears as a
dynamical variable, although this Lagrangian has some unsatisfactory 
features which we trace to peculiarities of the null limit. We also 
show that the M5-brane action is invariant under all (super)isometries 
of the background. 

\end{titlepage}
\newpage
\renewcommand{\thefootnote}{\arabic{footnote}}
\setcounter{footnote}{0} 

\section{Introduction}

The essential `ingredients' of M-theory that are additional to those of eleven
dimensional (D=11) supergravity are the supermembrane, or M2-brane, and the
M5-brane. The worldvolume action for the supermembrane has been known for more
than ten years \cite{BST}. In contrast, the full worldvolume action for the
M5-brane has been known for only a year or so and its implications are still
being explored. The M5-brane action is essentially one for an interacting
six-dimensional (2,0) supersymmetric gauge theory based on the (2,0)
antisymmetric tensor supermultiplet \cite{HST}. The self-duality of the
3-form field strength of this supermultiplet presents serious obstacles to the
construction of a six-dimensional Lorentz covariant action, some of
which are inevitable at the quantum level \cite{W1}. This is not a problem at
the level of field equations, however, although the self-duality constraint
involves non-linearities that would be hard to guess \cite{PS}. The full field
equations were found in superfield form in \cite{HS}. A covariant component
action involving an additional scalar gauge field was presented in
\cite{PST}, although it is restricted to backgrounds admitting a nowhere-null
vector field. This action was used in \cite{ST} to determine the 
central charge structure of the M5-brane supertranslation algebra in a vacuum 
background. Alternatively, an action can be constructed by relaxing the
requirement of six-dimensional Lorentz covariance to five-dimensional Lorentz
covariance \cite{APPS}. These various formulations of the M5-brane action are
now known to be equivalent \cite{HSW,BLNPST}. 

In this paper we shall present another formulation of the M5-brane action: the
Hamiltonian formulation. The Hamiltonian formulation of the M2-brane can be
found in \cite{CBST}. The space/time split implicit in the Hamiltonian
formulation has the advantage that the self-duality constraint is no problem
and, in fact, is reduced to a simple {\sl linear} constraint on phase space.
Also, the Hamiltonian formulation is the natural one for investigations of static
solutions that minimise the energy, and some of the results obtained here were
advertized and then used for this purpose in
\cite{GGT}.

The passage to the Hamiltonian formulation from the Lagrangian one is
simplest if one starts from the covariant action of \cite{PST} because much
of the required space/time split can be achieved by the choice of temporal
gauge for the auxiliary scalar gauge field of this action (the `PST'
field). We shall therefore begin by reviewing those elements of this
formulation that are essential to the subsequent steps. In doing so we take the
opportunity to show, following a similar recent demonstration for super
D-branes \cite{BT}, that the M5-brane action is invariant under all
(super)isometries of the D=11 supergravity background, provided that the 2-form
gauge potential is assigned an appropriate transformation. This observation
acquires importance in light of the recent construction of an interacting
conformal invariant antisymmetric tensor field theory via gauge fixing of the
bosonic sector of the M5-brane action in an $adS_7\times S^4$ background
\cite{CKKTV}. Specifically, it implies that the full M5-brane action in this
background is invariant under the full $OSp(6,2|4)$ isometry supergroup of the
$adS_7\times S^4$ solution \cite{PTV} of D=11 supergravity. This symmetry will
be realized `on the brane' as a non-linearly realized six-dimensional
superconformal invariance.

For all branes other than the M5-brane it is known that the tension can be
replaced by a dynamical p-form worldvolume gauge potential, leading to a
Lagrangian that is strictly invariant under background isometries (as against
invariant up to the addition of a total derivative) by virtue of appropriate
transformations of the new worldvolume field \cite{tension,BT}. It was suggested
in \cite{town} that this may not be possible for the M5-brane. On the other
hand, a version of the M5-brane action with dynamical tension was found in
\cite{CNS}, although it did not incorporate the self-duality constraint (which
had to be imposed separately). Thus, at present, the status of dynamical tension
in the M5-brane case is unclear and one of the aims of this paper is to
shed some light on this point. The Hamiltonian formulation provides the means to
do so; in this formulation it is a simple matter to elevate the tension to the
status of a dynamical variable. One can then pass to the corresponding
covariant Lagrangian formulation in which the tension is replaced by a 5-form
gauge potential. We find that this Lagrangian has some unsatisfactory 
features, which we trace to peculiarities of the tensionless limit. 

\section{The M5-brane action and its rigid symmetries}

Any solution of the field equations of D=11 supergravity is a consistent
background for the M5-brane. These backgrounds can be presented as tensors on
D=11 superspace, which is parameterised by the coordinates $Z^M = (X^m,
\Th^{\a})$. Specifically, we need a D=11 supervielbein $E_M{}^A$, which are the
coordinate basis components of the frame 1-forms $E^A = (E^a,E^\a)$, where $E^a$
and $E^\a$ are, respectively, a vector and a Majorana spinor of the D=11
Lorentz group.  We also need a superspace 3-form gauge potential $C^{(3)}$ and a
6-form gauge potential $C^{(6)}$. Their gauge transformations are
\bea
\delta C^{(3)} &=& d\Lambda^{(2)}\, , \nn
\delta C^{(6)} &=& d\Lambda^{(5)} - {1\over2} \Lambda^{(2)} R^{(4)}\, ,
\label{aonea}
\eea
where $\Lambda^{(2)}$ and $\Lambda^{(5)}$ are 2-form and 5-form parameters,
respectively,  and their gauge-invariant field strengths are
\bea
R^{(4)} &=& dC^{(3)}\, , \nn
R^{(7)} &=& dC^{(6)} +{1\over2} C^{(3)} R^{(4)}\, ,
\label{onea}
\eea
where the exterior product of forms is understood. The relative sign in the
definition of $R^{(7)}$ differs from (e.g.) \cite{ST} because we use here 
the convention that the exterior superspace derivative $d$ `acts from the
right', i.e.
\be
d(PQ) = PdQ + (-1)^q(dP)Q
\ee
for $p$-form $P$ and $q$-form $Q$. The on-shell superfield constraints of D=11
supergravity imply, {\sl inter alia}, that the bosonic component of
$R^{(7)}$ is the D=11 Hodge dual of the bosonic component of $R^{(4)}$. 

Before turning to the M5-brane itself, let us note that a Killing vector
superfield $\xi(Z)$ is one for which
\be
{\cal L}_\xi (E^a \otimes_s E^b)\eta_{ab} =0
\label{oneb}
\ee
where ${\cal L}_\xi$ denotes the Lie derivative with respect to $\xi$, and
$\eta$ is the D=11 Minkowski metric. This is the superfield version of the
Killing condition. By an `isometry' of the supergravity background we shall
mean a transformation generated by a Killing vector field for which,
additionally,
\be
{\cal L}_\xi R^{(4)} =0, \qquad {\cal L}_\xi R^{(7)} =0.
\label{onec}
\ee
It is convenient to summarize the action of the complete set of such Killing
vector superfields $\xi_\alpha$ by means of a BRST operator $s$, so that
\be
sZ^M = c^\alpha\xi_\alpha^M \equiv c^M \, ,\qquad 
sc^\alpha = {1\over2} c^\gamma c^\beta f_{\beta\gamma}{}^\alpha\, ,
\label{oned}
\ee
where $c^\alpha$ is a set of {\sl constant} BRST `fields' and
$f_{\beta\gamma}{}^\alpha$ are the structure constants of the Lie algebra of
Killing vector fields. It follows that
$s^2c^\alpha\equiv 0$, and that $s^2Z^M\equiv 0$ (and hence that the action of
$s^2$ on any superfield vanishes identically).  Thus, (\ref{onec}) reduces to
$sR^{(4)}=sR^{(7)}=0$, which implies that\footnote{The same symbols were used in
\cite{ST} with a related but not identical meaning. Here we follow the notation
of \cite{BT}.}
\be
sC^{(3)} = d\Delta_2\, , \qquad
sC^{(6)} = d\Delta_5 - {1\over2} \Delta_2 R^{(4)}\, 
\label{onee}
\ee
where $\Delta_p$ is a ghost-valued superspace $p$-form. 

Now, let $\sigma^i ~(i=0,1,\dots,5)$ be the worldvolume coordinates of the
fivebrane and $f$ a map from the worldvolume to superspace. We take the
worldvolume metric to be the pullback $g=f^*(E^a \otimes_s E^b)\eta_{ab}$. It
is manifestly invariant under isometries of the background. The worldvolume
fields include, in addition to $Z^M(\sigma)$, a 2-form worldvolume gauge
potential $A(\sigma)$ with `modified' 3-form field strength\footnote{Here we
adopt a normalization of the two-form potential $A$ that differs by a factor of
three from \cite{ST}.}
\be
\label{19}
H=dA-C^{(3)}\, ,
\ee
where $C^{(3)}$ is now to be understood as the pullback of the
corresponding superspace 3-form gauge potential. Clearly, $H$ is invariant under
isometries of the background provided that we choose
\be
sA= \Delta_2\, .
\ee
We shall also need to define\footnote{Our metric signature is `mostly plus'.}
\be
\tilde H^{ij} = {1\over 6\sqrt{gg^{lm}\partial_l a\partial_m a}}\, 
\varepsilon^{ijki'j'k'}(\partial_k a) H_{i'j'k'}
\ee
where $g = \det g_{ij}$ and $a(\sigma)$ is the `PST' scalar
gauge field. The 2-form $\tilde H$ is also invariant under isometries of the
background if we take $sa=0$. Having made the above definitions, and 
introducing the fivebrane tension $T$, we may
write down the M5-brane Lagrangian of \cite{PST}. This was originally 
written in the form 
\be\label{or}
L_{M5} = T\big[ L_{DBI} + \tilde H H + L_{WZ}\big]
\ee
where 
\be\label{DBI}
L_{DBI} = -\sqrt{-\det (g_{ij} + \tilde H_{ij})}\, ,
\ee
is a type of Dirac-Born-Infeld action, and
\be\label{hhtilde}
\tilde H H = {1\over 24 (\partial a)^2}\, (\partial_i a) 
\varepsilon^{ijki'j'k'}H_{i'j'k'}H_{jkl}g^{ll'}\partial_{l'} a\, .
\ee
The last, Wess-Zumino (WZ), term will be given below. 

Here we shall start from the equivalent Lagrangian
\be\label{equivor}
L = L_0 + TL_{WZ}\, .
\ee
where 
\be\label{Lzero}
L_0 = {1\over2v} \det (g_{ij} + \tilde H_{ij}) - T^2 v
+ T\tilde H H
\ee
with $v(\sigma)$ an independent worldvolume scalar density. If we take
$sv=0$, in addition to the previously assigned BRST transformations, then $L_0$
is clearly invariant under isometries of the background. The WZ term is
\bea
L_{WZ} &=& {1\over 6!} \varepsilon^{ijklmn}\big[ C^{(6)}_{ijklmn} + 
10 H_{ijk} C^{(3)}_{lmn}\big] \nn
&=& \star \big(C^{(6)} + {1\over2} H C^{(3)}\big)
\eea
where $\star$ is the worldvolume Hodge dual, and $C^{(6)}$ and $C^{(3)}$ are 
now understood to be the pullbacks to the worldvolume of the corresponding
superspace forms. Now,
\be
s \big(C^{(6)} + {1\over2} H C^{(3)}\big) = 
d\big(\Delta_5 +{1\over2} H\Delta_2 \big)
\ee
where the ghost-valued forms $\Delta_p$ are also to be understood now as
pullbacks to the worldvolume of the corresponding superspace $p$-forms 
(recall that $d$ `acts from the right').

Thus, $L_{WZ}$ is invariant under isometries of the background up to a possible
total derivative. This is sufficient to ensure the existence of a conserved
worldvolume current associated  with each isometry of the background, although
the total derivative term, and $\Delta_2$ if non-zero, can lead to central
terms in the algebra of isometries, as discussed in \cite{ST} for the case of a
vacuum D=11 background.

\section{The Hamiltonian formulation}

As shown in \cite{PST}, the scalar field $a$ is subject to a gauge
transformation that allows the choice of `temporal gauge' $a=\sigma^0\equiv t$.
This choice breaks the $SO(1,5)$ Lorentz group to the $SO(5)$ rotation group, 
but this is in any case an expected feature of the Hamiltonian formulation. 
Our goal is to separate in the Lagrangian $L$ all terms involving time derivatives of the worldvolume fields. Let us set $\sigma^i=(t,\sigma^\ua)$, 
($\ua=1,\dots 5$). Then, in the $a=t$ gauge
we have
\be
\tilde H H
= {1\over 24} \varepsilon^{\ua\ub\uc\ud\ue} H_{\uc\ud\ue} H_{\ua\ub 0} - V_\ua
{}^5\!g^{\ua\ub}g_{\ub0}
\ee
where ${}^5\!g^{\ua\ub}$ is the inverse of $g_{\ua\ub}$ (rather than the
space/space components of $g^{ij}$) and
\be
V_\uf = {1\over 24} \varepsilon^{\ua\ub\uc\ud\ue}H_{\uc\ud\ue} H_{\ua\ub\uf}\, .
\ee
In addition, the only non-vanishing component of $\tilde H^{ij}$ in the gauge
$a=t$ is 
\be
\tilde H^{ab} = {1\over 6\, \sqrt{{}^5\!g}}\, \varepsilon^{abcde} H_{cde}
\ee
where ${}^5\!g$ is the determinant of the worldspace 5-metric $g_{\ua\ub}$.
It follows that
\be
\det (g_{ij} + \tilde H_{ij}) = 
(g_{00} - g_{0\ua}{}^5\!g^{\ua\ub} g_{0\ub}) \det {}^5\!(g+ \tilde H)
\ee
where 
\be
\tilde H_{\ua\ub} = g_{\ua\uc}g_{\ub\ud}\tilde H^{\uc\ud} \, ,\qquad
\det {}^5\!(g+ \tilde H) = \det (g_{\ua\ub} + \tilde H_{\ua\ub} )\, .
\ee
If we now define
\be\label{redef}
\lambda = v/\det {}^5\!(g+ \tilde H)
\ee
we find that
\be
L_0 = {1\over2\lambda} (g_{00} - g_{0\ua}{}^5\!g^{\ua\ub} g_{0\ub}) 
-T V^{\ua} g_{0\ub} - {1\over 2}\lambda T^2 \det {}^5\!(g+ \tilde H) + 
{T\over 24} \varepsilon^{\ua\ub\uc\ud\ue} H_{\uc\ud\ue} H_{\ua\ub 0}\, 
\label{ellzero}
\ee
where $V^\ua = {}^5\!g^{\ua\ub}V_\ub$. 

We must now make the space/time split in the WZ term. We have
\be
L_{WZ} = \dot Z^M \big[{\cal C}_M - 
{1\over24} \varepsilon^{\ua\ub\uc\ud\ue} \dot Z^M 
C^{(3)}_{Mab}H_{\uc\ud\ue}\big] + {1\over
24}\varepsilon^{\ua\ub\uc\ud\ue}C^{(3)}_{\uc\ud\ue} H_{\ua\ub 0}
\ee
Here we have introduced the worldspace scalar
\be
{\cal C}_M = * i_M C^{(6)}\, ,
\ee
where $i_M$ indicates contraction with the vector 
field $\partial/\partial Z^M$ (so that $i_M C^{(6)}$ is a 5-form, which we 
restrict to the worldspace)
and $*$ is the {\sl worldspace} Hodge dual. 

We have now implicitly arrived at a space/time split form of the total M5-brane
Lagrangian $L= L_0 + TL_{WZ}$. It will be convenient to write this result as
\be
L= L_1 + L_2
\ee
where $L_1$ includes all but the last term (with the time component of $H$) in
(\ref{ellzero}) and $L_2$ is the rest. An equivalent form for $L_1$ is
\be
L_1 = \tilde P \cdot \Pi_t - s^\ua (\tilde P \cdot \Pi_\ua + TV_\ua)
-{1\over2}\lambda\big[ (\tilde P + T V^\ua\Pi_\ua)^2 + T^2 \det{}^5\!(g +
\tilde H)\big]\, ,
\ee
where $\tilde P^a$ and $s^\ua$ are new independent variables
and we have set
\be
(\Pi_i)^a \equiv E_i{}^a = \partial_iZ^M E_M{}^a\, .
\ee
Recall that the indices $a,b,\dots$ denote D=11 Lorentz vectors, which are
contracted with the Minkowski metric $\eta$. The equivalence may be established
by successive elimination of $\tilde P$ and $s^\ua$. 

The Lagrangian $L_2$ is
\be
L_2 = T \dot Z^M \big[{\cal C}_M - {1\over 24} \varepsilon^{\ua\ub\uc\ud\ue}
C^{(3)}_{M\ua\ub} H_{\uc\ud\ue} \big]
 + {1\over24}T \varepsilon^{\ua\ub\uc\ud\ue} H_{\ua\ub 0} (dA)_{\uc\ud\ue}\, .
\ee
Now,
\be
H_{\ua\ub 0} = \dot A_{\ua\ub} + 2\partial_{[\ua}A_{\ub]0} - \dot Z^M
C^{(3)}_{M\ua\ub}
\ee
so, omitting a total derivative term, we have
\be
L_2 = T \dot Z^M \hat{\cal C}_M + 
{1\over8}T\varepsilon^{\ua\ub\uc\ud\ue} \dot A_{\ua\ub}\partial_\uc A_{\ud\ue}
\ee
where 
\bea
\hat{\cal C}_M &=& {\cal C}_M - {1\over24} \varepsilon^{\ua\ub\uc\ud\ue}
C^{(3)}_{M\ua\ub}[C^{(3)}_{\uc\ud\ue} + 2H_{\uc\ud\ue}] \nn
&=& *\big[i_MC^{(6)} -{1\over2} i_MC^{(3)}(C^{(3)} + 2H)\big]\, .
\eea
We can further rewrite this as
\be
L_2 = T \dot Z^M \hat{\cal C}_M + {1\over2}\Pi^{\ua\ub} \dot A_{\ua\ub}
\ee
where
\be
\Pi^{\ua\ub} ={1\over4}T\varepsilon^{\ua\ub\uc\ud\ue} 
\partial_\uc A_{\ud\ue}\, .
\ee

Putting all these results together we can write the M5-brane lagrangian in the
form
\be
\label{halfway}
L= \tilde P \cdot \Pi_t + {1\over2}\Pi^{\ua\ub} \dot A_{\ua\ub} 
-\lambda {\cal H} -s^\ua \tilde {\cal H}_\ua  + \sigma_{\ua\ub} 
{\cal K}^{\ua\ub}  + T\dot Z^M \hat {\cal C}_M
\ee
where
\bea
{\cal H} &=& {1\over2}\big[(\tilde P + T V^\ua\Pi_\ua)^2 + T^2
\det{}^5\!(g + \tilde H)\big] \nn
\tilde {\cal H}_\ua &=& (\tilde P \cdot \Pi_\ua + TV_\ua) \nn
{\cal K}^{\ua\ub} &=& \Pi^{\ua\ub}-{1\over4}T\varepsilon^{\ua\ub\uc\ud\ue} 
\partial_\uc A_{\ud\ue}
\eea

We have now arrived at a `half way house' on the way to the fully canonical
phase space form of the M5-brane Lagrangian, and it is convenient to pause here
to assess the situation. Note, in particular, that the constraint ${\cal
K}^{\ua\ub}=0$ can be written in differential form notation on worldspace as
\be\label{sdconstraint}
\Pi = {1\over2}T * (dA)
\ee
which implies the Gauss law constraint $d*\Pi =0$. In other words, the
constraint ${\cal K}^{\ua\ub}=0$ ensures that the Bianchi identity for $dA$
implies the Gauss law, as expected in a self-dual antisymmetric tensor
field theory. Thus {\sl the self-duality constraint that is so problematic in 
the Lagrangian reduces to a set of simple linear constraints on the phase space
pair} $(\Pi,A)$. However, these constraints are second class, in Dirac's
terminology, and this leads to the problems upon quantization. One must either
solve the constraints (which in general leads to non-locality) or convert them
into first class ones by the addition of auxiliary variables. As explained in
\cite{BK}, the latter approach leads to the infinite-field formulation of chiral
antisymmetric tensors of \cite{DH,Berk}. The reason that we find these bosonic
second-class constraints is that we fixed the gauge for the PST field $a$ in the
passage to the Hamiltonian formulation. Had we not fixed this gauge invariance
we would have found only first class bosonic constraints with an additional
constraint associated with the additional gauge invariance, as found for the
free chiral D=6 2-form field in \cite{PSTb}. 

From (\ref{halfway}) we see that the momentum $P_M$ conjugate to $Z^M$ is given
by
\be
\label{momentum}
P_M = E_M{}^a \tilde P_a + T \hat{\cal C}_M
\ee
Solving for $\tilde P$ we have
\be
\label{solvep}
\tilde P_a = E_a{}^M P_M - TE_a{}^M\hat{\cal C}_M\, .
\ee
The remaining information contained in (\ref{momentum}) is the
fermionic constraint
\be
P_\mu - E_\mu{}^aE_a{}^M P_M = T\big[ \hat{\cal C}_\mu -
E_\mu{}^aE_a{}^M\hat{\cal C}_M\big]
\ee
which is equivalent to
\be
E_\mu{}^\alpha E_\alpha^M (P_M-T\hat{\cal C}_M) =0
\ee
Since $E_\mu{}^\alpha$ is invertible, this constraint is equivalent to
${\cal S}_\alpha =0$, where
\be
{\cal S}_\alpha = E_\alpha{}^M (P_M -T\hat{\cal C}_M)\, .
\ee
This constraint can be imposed by a new spinorial Lagrange multiplier
$\zeta^\alpha$. It can also be used to simplify the constraint imposed by
$s^\ua$ since
\be
\tilde{\cal H}_\ua = {\cal H}_\ua + 
\partial_\ua Z^ME_M{}^\alpha {\cal S}_\alpha
\ee
where
\be
{\cal H}_\ua = \partial_\ua Z^M P_M + 
T(V_\ua -\partial_\ua Z^M \hat{\cal C}_M)\, .
\ee

We thus arrive at the M5-brane Lagrangian in fully canonical
form
\be
\label{phase}
L= \dot Z^M P_M + {1\over2}\Pi^{\ua\ub} \dot A_{\ua\ub} 
-\lambda {\cal H} -s^\ua {\cal H}_\ua  + \sigma_{\ua\ub} {\cal K}^{\ua\ub} 
- \zeta^\alpha {\cal S}_\alpha
\ee
where
\bea
{\cal H} &=& {1\over2} \big[{\cal P}^2 + T^2 \det{}^5\!(g + \tilde H)\big] \nn
{\cal H}_\ua &=& \partial_\ua Z^M P_M + T(V_\ua -\hat{\cal C}_\ua)\nn
{\cal K}^{\ua\ub} &=& \Pi^{\ua\ub}-{1\over4}T\varepsilon^{\ua\ub\uc\ud\ue} 
\partial_\uc A_{\ud\ue}\nn 
{\cal S}_\alpha &=& E_\alpha{}^M (P_M -T\hat{\cal C}_M)
\eea
with
\be\label{phasethree}
{\cal P}_a = E_a{}^MP_M + 
T(V^\ua\partial_\ua Z^M E_M{}^b\eta_{ba} - \hat{\cal C}_a)\, .
\ee
The $\kappa$-symmetry of the M5-brane action is now reflected (for backgrounds
allowing $\kappa$-symmetry) in the fact that the fermionic constraints ${\cal
S}$ are half first-class and half second-class.

\section{Dynamical tension and the null limit}
 
We turn now to the issue of dynamical M5-brane tension.
This can be achieved in the Hamiltonian formulation 
by declaring $T$ to be an independent variable and then adding to the
Lagrangian ({\ref{phase}) the new term
\be
L' = \dot\phi T - u^\ua \partial_\ua T
\ee
where $\phi$ is a variable canonically conjugate to $T$, and $u^\ua$ is a
Lagrange multiplier for a new (first-class) constraint. This is the 
phase space form of the action for a 5-form gauge potential. In fact, 
eliminating momenta to
return to the Lagrangian form we find an equivalent Lagrangian for the
original worldvolume fields $(Z^M, A)$ together with a new 5-form gauge 
potential $A^{(5)}$. This Lagrangian can be shown to be the restriction to
$a=t$ of a new six-dimensional Lorentz covariant Lagrangian that depends
additionally on the PST field $a$. This covariant Lagrangian is
\be\label{newmfive}
L= {1\over v}\big[L_{DBI}^2 - (\star G)^2\big]
\ee
where $L_{DBI}$ is the Lagrangian given in (\ref{DBI}) and
and $G$ is the `modified' 6-form field strength
\be\label{dynamict}
G= dA^{(5)} - L_{WZ} - \tilde H H\, .
\ee
The $\tilde H H$ term is the same as the one in (\ref{hhtilde}); written 
in differential form notation it is
\be
\tilde H H \equiv {1\over4(\partial a)^2}da\wedge H \wedge i_{\partial a}H \,.
\ee
The reason that this term appears in $G$ is that the original
M5-brane Lagrangian changes by a total derivative under the transformation 
\be
\delta A = da\wedge \varphi^1(\sigma)
\ee
with local parameter $\varphi^1$ \cite{PST} (note that $L_{DBI}$ is
invariant under this transformation). The $\tilde HH$ term therefore 
behaves like
a WZ term with respect to this transformation and its non-invariance is
compensated in $G$ by an appropriate transformation of $A^{(5)}$. The 
Lagrangian (\ref{newmfive}) is the M5-brane analogue of the super D-brane
Lagrangian of \cite{BT}.  

Since we have now reintroduced the PST field we expect the Lagrangian
(\ref{newmfive}) to be invariant under a gauge transformation that will
allow the PST field $a$ to be eliminated by a choice of gauge (e.g. the
`temporal' gauge $a=t$). This is not guaranteed by the construction (so far we
know only that we recover a Lagrangian equivalent to the original on setting
$a=t$) so it must be checked. We expect this gauge invariance to take the form
\be\label{vara}
\delta a = \varphi(\sigma), \qquad \delta A = \varphi(\sigma) {\cal L}^2\, ,
\ee
together with some variation $\delta v$ of the Lagrange multiplier $v$;
the precise form of ${\cal L}^2$ can be found in \cite{PST}. To find the
variation of $v$, consider a general variation of 
(\ref{newmfive}). This has the form
\bea\label{vary}
\delta L &=&{2\over v}[L_{DBI}\delta L_{DBI}+ 
\star G \delta (\tilde H H+L_{WZ})
-\star G d(\delta A^5)] -{1\over v^2}[(L_{DBI})^2- (\star G)^2]\delta v \nn
&=&{2\over v}\big\{(L_{DBI}-\star G)\delta L_{DBI}
+\star G[\delta (L_{M5}) -d\delta A^5]\big\} \nn
&&-{1\over v^2}(L_{DBI}-\star G)(L_{DBI}+\star G)\delta v\, ,
\eea
where $L_{M5}= L_{DBI}+\tilde H H + L_{WZ}$
is the original Lagrangian of \cite{PST}. We know, for {\sl any} of the
symmetry transformations of this action that $\delta L_{M5}=d\Lambda$
for some function $\Lambda$. This total derivative can be cancelled by a choice
of $\delta A^{(5)}$. We are therefore left with
$$
\delta L={2\over v}(L_{DBI}- \star G)\delta L_{DBI}
-{1\over v^2}(L_{DBI}- \star G)(L_{DBI}+\star G)\delta v\, ,
$$
which vanishes if 
\be\label{v}
\delta v={{2v\delta L_{DBI}}\over{(L_{DBI}+\star G)}}\,.
\ee

It would seem from this result that we have now found a covariant M5-brane
action in which the tension is replaced by a 5-form gauge potential, as has
now been achieved for most other branes. There is a difficulty, however. The
denominator of the expression \p{v} for $\delta v$ can vanish on the mass-shell.
The mass-shell constraint (imposed by the Lagrange multiplier $v$) is
\be\label{ve}
0= (L_{DBI})^2-(\star G)^2 = (L_{DBI}- \star G)(L_{DBI}+ \star G)
\ee
which implies that $\star G= \pm L_{DBI}$. If we choose the minus sign then the
variation $\delta v$ is singular. Note that this problem occurs {\sl only} for
the PST gauge transformation.  For $\kappa$-transformations, the
denominator in \p{v} is cancelled  by a factor appearing in $\delta_\kappa
L_{DBI}$ (as occurs for super D-branes \cite{BT}). For rigid symmetries
associated with isometries of the background we have $\delta L_{DBI}=0$ and
hence $\delta v=0$. There is therefore no problem for D-branes or any
other brane for which a Lorentz covariant action is possible without the
introduction of a PST gauge scalar, but there is a problem for the M5-brane. If
we demand that the PST gauge variation of $v$ be non-singular then we are
required to choose the solution
$$
\star G = L_{DBI}
$$
of the mass-shell constraint. This choice corresponds to the self--duality
equations produced by the Lagrangian \p{or}. We conclude that the Lagrangian
(\ref{newmfive}) is classically  equivalent to the original M5-brane Lagrangian
(\ref{or}) only if the solutions of the equations of motion of the former are
restricted in the way just described. Thus, we have found a version of the
covariant M5-brane action of \cite{PST} in which the tension appears as a
dynamical variable, but it is not completely satisfactory in that it is 
necessary to suplement the equations of motion that follow from the action
with additional information. This is equally a defect of the proposal of
\cite{CNS} for dynamical M5-brane tension.

The unsatisfactory feature of the action \p{newmfive} that we have just
explained has implications for the null, i.e. tensionless, limit. If we take
this limit in \p{newmfive}, by setting $G=0$, then the variation \p{v} becomes
singular on the mass shell since the mass-shell constraint is now $L_{DBI}=0$.
This indicates that the zero tension limit can be taken consistently only if one
simultaneously sets to zero the worldvolume 2-form field $A$, because in this
case the gauge transformations \p{vara} disappear. In this case we recover the
standard null super-fivebrane Lagrangian (i.e. without the BI or WZ terms). This
is effectively the $T=0$ case of \p{phase} because when $T=0$  the constraint
(\ref{sdconstraint}) implies that $\Pi=0$ and the conjugate 2-form $A$ then
drops out of the Lagrangian. 

Actually, some of these features of the action \p{newmfive} are already present
in the Lagrangian \p{equivor}. Under \p{vara} the variation of the Lagrange
multiplier $v$ should take the form
\be
\delta v= \big[T-v^{-1}L_{DBI}\big]^{-1}
\big[\delta (\tilde H H + L_{WZ})-d\Lambda\big]\, 
\ee
where $d\Lambda$ is the variation of the original Lagrangian $L_{M5}$.
The requirement that this variation be non-singular now singles out one of two
possible solutions for $v$ (namely $v=-TL_{DBI}$) and the $T\rightarrow 0$ 
limit can be taken only if $H$ is set to zero. In the passage to the
Hamiltonian formulation we effectively made this choice by the redefinition
\p{redef} of the Lagrange multiplier, which explains why $H$ drops out of
the phase-space action when $T=0$. The problem in the Lagrangian formulation is
not solved by gauge fixing the PST invariance because after the gauge fixing the
Lagrangian is still invariant (up to a total derivative) under a combination of
a PST gauge transformation with a (now non--manifest) worldvolume diffeomorphism
that preserves the gauge condition. This is an essential symmetry of the 
non-covariant formulation of the five--brane action of \cite{APPS}. Moreover,
these non-manifest worldvolume diffeomorphisms will be preserved when rewriting
the five--brane Lagrangian in any other form; it is this fact that leads to the
transformation properties of the Lagrange multiplier discussed above.

The conclusion seems to be that if we consider the PST gauge invariance and/or
(non-manifest) covariance of the five--brane actions as essential properties
then the zero-tension limit leads to an ordinary null super-fivebrane, and so
does not commute with double dimensional reduction of the five--brane to a dual
IIA D4--brane. If we sacrifice these symmetries and pass to an intrinsically
non-covariant formulation of the M5-brane, then the problem with the variation
of Lagrange multipliers does not arise. In this case the zero tension limit may
commute with the double dimensional reduction, although probably only for those
cases that correspond to gauge fixing $a(\sigma)$ to be the spatial coordinate 
of the compactified dimension of the five--brane. Note also that in this case 
the 5-brane action will yield equations of motion for  either a self-dual or an
anti-selfdual worldvolume field, depending on which solution of \p{ve} is
chosen.

\section{Discussion}

We have presented in (\ref{phase}-\ref{phasethree}) a Hamiltonian form of the
M5-brane Lagrangian. This result contains various subcases of special interest.
For example, the bosonic Lagrangian in a vacuum background is
\be
L_{bos}= P \cdot \dot X  + {1\over2}\Pi^{\ua\ub} \dot A_{\ua\ub} 
-\lambda {\cal H} -s^\ua {\cal H}_\ua  + \sigma_{\ua\ub} {\cal K}^{\ua\ub} 
\ee
where 
\bea
{\cal H} &=& {1\over2}\big[ (P+ TV^\ua\partial_\ua X)^2 + T^2 \det {}^5\!
(g+\tilde H)\big] \nn 
{\cal H}_\ua &=& \partial_\ua X \cdot P + TV_\ua \nn
{\cal K}^{\ua\ub} &=& \Pi^{\ua\ub}-{1\over4}T\varepsilon^{\ua\ub\uc\ud\ue} 
\partial_\uc A_{\ud\ue}
\eea
This is the result advertized in \cite{GGT} (with $T=1$ and different factors
arising from slightly different conventions). 

Another special case is the null M5-brane Lagrangian. Setting $T=0$ we find
the Lagrangian
\be\label{null}
L\big|_{T=0} = \dot Z^M P_M -\lambda {\cal H}
-s^\ua {\cal H}_\ua - \zeta^\alpha {\cal S}_\alpha
\ee
where
\be
{\cal H}={1\over2}\eta^{ab}E_a{}^M E_b{}^N P_N P_M\, , \qquad
{\cal H}_\ua = \partial_\ua Z^M P_M\, ,\qquad 
{\cal S}_\alpha= E_\alpha{}^M P_M\, . 
\ee
A curious feature of the null limit of the M5-brane is that the 2-form gauge
potential disappears. As we have seen, this can be traced to the requirement
that the tensionless limit be consistent with the  PST gauge invariance.  
This feature, and the related inadequacies of the M5-brane action with dynamical
tension, may well be a reflection of the absence of a compelling physical
argument in favour of the promotion of the M5-brane tension to a dynamical
variable. The basic reason that one needs to elevate a brane tension to the
status of a dynamical variable is to accomodate the possibility of it ending on
another brane, but the M5-brane cannot have a boundary on another brane. This is
the argument against dynamical M5-brane tension given in \cite{town}. 

This argument can be restated in terms of worldvolume domain
walls. These can occur as solutions to the brane equations of motion only if
discontinuities are allowed in the tension \cite{BPvS}, which is possible 
only if
the tension is replaced by a p-form gauge potential. Consider now the NS5A and
NS5B branes. The worldvolume solitons on these branes can be interpreted as
``little d-branes''. However, in this approach the little d-4-brane, or domain
wall, is a iib solution of the  NS5B brane equations. The systematics is as
follows \cite{pap}:
\vskip 0.3cm
\noindent d-0-brane: iib (needs $c^{(1)}$)  \hfill\break 
\noindent d-1-brane: iia (needs $c^{(2)+}$) \hfill\break 
\noindent d-2-brane: iib (needs $c^{(3)}=$ dual of $c^{(1)}$) \hfill\break 
\noindent d-3-brane: iia (needs $c^{(4)}=$ dual of $c^{(0)}$) \hfill\break 
\noindent d-4-brane: iib (needs $c^{(5)}=$ dual of $T$) \hfill\break
\vskip 0.3cm 
\noindent
where $c^{(r)}$ is a D-brane worldvolume $r$-form gauge potential.
Therefore one should replace the tension $T$ of the NS5B brane
by a dynamical variable but there is no need to do so for the tension of the 
NS5A brane. Since the M-theory origin of the NS5A brane is the M5-brane,
it follows that neither there is any need to replace the M5-brane tension by a
dynamical variable. Alternatively, one may note that the intersection of an
M5-brane with any other M-brane is always over a 1-brane, 3-brane or 5-brane but
never over a 4-brane. Therefore, there is never a need to construct a domain
wall solution to the M5-brane equations of motion and, correspondingly, there is
no need to replace the M5-brane tension by a dynamical variable.

\bigskip
\noindent
{\bf Acknowledgements.} The authors wish to thank Jerome Gauntlett, Joaquim
Gomis and Mario Tonin for helpful discussions.  Work of D.S. was partially
supported by research grants of the Ministry of Science and Technology of
Ukraine and the INTAS Grants N 93--127--ext. and N 93--0308.

\end{document}